\newcommand{\be}{\begin{equation}}
        \newcommand{\ee}{\end{equation}}
        \newcommand{\ba}{\begin{eqnarray}}
        \newcommand{\ea}{\end{eqnarray}}
        \newcommand{\ban}{\begin{eqnarray*}}
        \newcommand{\ean}{\end{eqnarray*}}
        \newcommand{\nl}{\nonumber \\} 
\newcommand{\R}{{\mathbb R}}  
  
\newcommand{\Z}{{\mathbb Z}}  
\newcommand{\N}{{\mathbb N}}  
	
\newtheorem{theorem}{Theorem}

\documentclass[11pt]{article}
\usepackage {graphicx,latexsym}
\usepackage{amsfonts}

\begin{document}

\title{
Loop quantization as a continuum limit}
\author{
Elisa Manrique$^{1,2}$
\footnote{e-mail: \ttfamily elisa@ifm.umich.mx}, 
Robert Oeckl$^1$
\footnote{e-mail: \ttfamily oeckl@matmor.unam.mx}, \\ 
Axel Weber$^2$ 
\footnote{e-mail: \ttfamily axel@ifm.umich.mx}
and Jos\'e A. Zapata$^1$
\footnote{e-mail: \ttfamily zapata@matmor.unam.mx} 
\\
{\it $^1$Instituto de Matem\'aticas UNAM} \\ 
{\it A.P. 61-3, Morelia Mich. 58090, M\'exico}\\
{\it $^2$Instituto de F\'{\i}sica y Matem\'{a}ticas,}\\ 
{\it Universidad Michoacana de San Nicol\'{a}s de Hidalgo,}\\
{\it Edif. C-3, C. U., Morelia, Mich.  58040, M\'{e}xico}
}

\date{}
\maketitle

\begin{abstract} 
We present an implementation of Wilson's renormalization group 
and a continuum limit tailored for loop quantization. 
The dynamics of loop quantized theories is constructed as a continuum limit of dynamics of effective theories. 
After presenting the general formalism we show as first explicit example the 2d Ising field theory. 
It is an interacting relativistic quantum field theory with local degrees of freedom quantized by loop quantization techniques. 
\end{abstract}

\section{Introduction and summary of results}

Loop Quantization was originally motivated by its application to 
theories free of a background metric like gravity and gravity coupled to matter. 
However, an extended version of the formalism 
sometimes referred to as a ``polymer representation" 
has been developed for  gauge theories with compact gauge group, sigma models with compact group and scalar fields and it is 
equally applicable to theories free of a background metric and to theories 
that use a background metric. 
Much of the interest in this extended formalism is 
the possibility of comparing with well established physics. 
The kinematics of 
such quantum field theories is very well understood. 
They are characterized by a space of generalized ``polymer-like" field configurations (connections in the 
case of gauge theories) $\bar{\cal A}_M$ that has been extensively studied. 
In this work we will loosely refer to theories in the continuum whose kinematical basis is a ``polymer representation" on the space 
$\bar{\cal A}_M$ as loop quantized theories. 
In contrast with its kinematical side, the dynamics of quantum theories constructed by Loop Quantization is regarded as well understood only for topological theories \cite{3dGrav} 
and for two dimensional Yang-Mills theories \cite{2dYM}. 

It has been argued that 
it would be natural to construct the dynamics of loop quantized theories as a continuum limit of the dynamics of effective theories following Wilson's renormalization group ideas, as it is done in lattice gauge theories \cite{LQinLGT}. 
However, the concept of scale that lies at the center of the renormalization group is background metric dependent. 

In our work we overcome this obstacle postulating an ``extended notion of scale;" then we implement Wilson's Renormalization Group and the continuum limit based on this notion. 
When such a continuum limit exists it defines the dynamics of a 
loop quantized theory. 
If this ``extended notion of scale" is used in metric dependent theories, it can be reduced to standard regular lattices. 
{\em Structures of the type presented in this letter should be able to bring 
a lot of the work on metric theories done by ordinary Lattice Gauge Theory (analytical and numerical) to the extended Loop Quantization framework.} 

In discrete approaches to quantum gravity the intention to implement 
Wilson's renormalization group goes back to Regge calculus \cite{RGinRegge} and reappears in the context of spin foam models \cite{RGsfFotini,RGsfRobert}. This work was suspected to be related to loop quantum gravity, but an explicit relation between the discrete models and the continuum of loop quantum gravity was missing. We fill this gap in this work and in \cite{GeoC-flat}; in particular the ``extended notion of scale" mentioned above places a cut-off in the continuum of loop quantization that yields discrete models of the type used in \cite{RGsfRobert}. 

We conclude with an explicit example; using our implementation of Wilson's renormalization group and our continuum limit we construct the dynamics of a loop quantized quantum field theory. 
The example is a very well studied quantum field theory: 2d Ising field theory. It is an interacting relativistic quantum field theory with local degrees of freedom quantized by loop quantization techniques. 

The paper is organized as follows. In section \ref{prelim} we review basic notions and notation of cellular decompositions and loop quantization. Our implementation of Wilson's renormalization group and the continuum limit are presented in section \ref{RG}. The example of the 2d Ising field theory is presented in the closing section.

\section{Basic notions} \label{prelim}

{\em Cellular decompositions}.
We start with a review of a few preliminary notions and notation on cellular decompositions of manifolds. 
In this work by a {\em cellular decomposition} $C$ of a manifold $M$ we mean
a presentation of it as a locally finite union of disjoint cells 
\[
M= \cup_{c_{\alpha \in C}} c_{\alpha} \quad , 
\quad
 c_{\alpha} \cap  c_{\beta} = \emptyset \hbox{ if } \alpha \neq \beta .
\]
Each cell $c_{\alpha}$ is the embedding of an open convex polyhedron of dimension between zero and $\dim M$. 
A typical example of a cellular decomposition is a triangulation of the sphere whose cells are: four triangles, six edges and four vertices. 

We will be concerned with the calculation of $n$-point functions. Thus, we will have a manifold $M$ with a set of $n$ marked points $\{ p_1, \ldots , p_n \}$. A {\em generic} cellular decomposition of a manifold $M$ with $n$ marked points is a cellular decomposition of $M$ such that each of the marked points is contained in a cell of the same dimension as $M$. 

The families of cellular decompositions that we use enjoy three properties
which let them play the role of  ``cut-off scales": 
\begin{enumerate}
\item[($i$)] {\em Partial order relation}. Two cellular decompositions are related $C_1 \leq C_2$ 
if any cell in the coarser decomposition $C_1$ is a finite union of cells of the finer decomposition $C_2$. 
\item[($ii$)] {\em Common refinement}. Given any two cellular decompositions $C_1, C_2$ there is a common refinement; $C_3$ 
such that $C_1 \leq C_3$ and $C_2 \leq C_3$. This property makes the family of cellular decompositions a partially ordered and directed set. (Directed towards refinement.)
\item[($iii$)] {\em Infinite refinement}. Given any open set $U$ of $M$ there is a fine enough cellular decomposition $C_0$ in the family with a cell $c_{\alpha} \in C_0$ of dimension $\dim(c_{\alpha}) = \dim(M)$ that is 
completely contained in the open set, $c_{\alpha} \subset U$. 
\end{enumerate}

There are many families of cellular decompositions enjoying these properties. 

For a given cellular decomposition $C$, the set of its cells will be denoted by $L(C)$; 
if $c_\alpha$ is a cell in $C$ we will call the corresponding element of $L(C)$ simply by its label $\alpha \in L(C)$. 
The elements of this abstract set can be thought of as points, but 
are not a priori related to specific points in $M$. 
However, the structure of the cellular decomposition has a preferred type of 
maps from $L(C)$ to $M$ that we will call {\em representative embeddings}. 
These maps  
${\rm Emb}_{L(C)}: L(C) \to M$ must obey  
${\rm Emb}_{L(C)} (\alpha) \in c_\alpha \subset M$ for every cell of the cellular decomposition. Clearly specifying one such representative embedding implies a choice. 

Given two cellular decompositions $C_1 \leq C_2$ there is a natural map 
$r_{2,1}: L(C_2) \to L(C_1)$ which sets $r_{2,1} (\alpha^{(2)}) = \alpha^{(1)}$ if and only if 
$c_{\alpha^{(2)}} \subset c_{\alpha^{(1)}}$ ($c_{\alpha^{(2)}}$ is a cell in $C_2$ and 
$c_{\alpha^{(1)}}$ is a cell in $C_1$). 
Again one can embed $L(C_1)$ into $L(C_2)$ at will, but we call an embedding 
${\rm Emb}_{1,2}: L(C_1) \to L(C_2)$ 
representative if $r_{2,1} \circ {\rm Emb}_{1,2} = {\rm id}$. 
With this we close our review of preliminary notions of cellular decompositions.

{\em Kinematics of the continuum: kinematics of loop quantized Euclidean Quantum Field Theories}. 
Here we will work in the Euclidean (imaginary time) description of Quantum Field Theory, but a similar structure exists also in a Hamiltonian description. In this work we will give the general prescription for spin systems, sigma models and scalar fields; for gauge systems the different aspects are treated carefully in \cite{GeoC-flat}.

The space of Euclidean histories will  be denoted by $\bar{\cal A}_M$ and its elements $s \in  \bar{\cal A}_M$ assign an element of a compact group%
\footnote{
The loop quantization (polymer representation) of the scalar field is based on a compact group. The Bohr compactification of $\R$ \cite{loopscalar}. 
} 
to any point of spacetime $s(p) \in {\rm G}$ without any continuity requirement. Thus, the algebra of fundamental observables is considered to be the so called ``cylindrical functions" $f \in {\rm Cyl}(\bar{\cal A}_M)$ which is composed of functions that depend on the histories restricted to finitely many points of spacetime, $f=f(\{ s(p_i)\})$. A very important particular case is the product of functions that depend on a single point of spacetime; the expectation value of such products are the familiar $n$-point functions. 

{\em Dynamics of the continuum: physical measure}. A physical measure $\mu_M$ on $ \bar{\cal A}_M$ lets us calculate expectation values and correlations among physical observables; $\langle f \rangle = \int_{\bar{\cal A}_M} f d\mu_M$ gives a precise meaning to expressions of the type 
$\langle f \rangle = \hbox{``} \frac{1}{Z} \int {\cal D \phi} \exp(-S(\phi)) f(\phi) \hbox{"}$. We stress that this measure encodes the dynamics of the theory in contrast to the auxiliary measure used to define the inner product in the usual kinematical Hilbert space of canonical loop quantized theories. The construction of physical measures is the primary goal of our work.

\section{Implementation of Wilson's renormalization group and the continuum limit}
\label{RG}

We define a space of {\em effective Euclidean histories at scale $C$} as the space of $C$-constant  field configurations, ${\cal A}_C$. By definition  $s \in {\cal A}_C \subset \bar{\cal A}_M$ if and only if for any $p, q \in M$ contained in the same cell of $C$ $s(p) = s(q)$. Clearly the space of $C$-constant functions on $M$ is in one to one correspondence with the space of functions on $L(C)$. 
Thus, there is a natural identification between ${\cal A}_C$ and the space 
${\cal A}_{L(C)}$ consisting of functions from $L(C)$ to ${\rm G}$. 
The algebra of effective observables at scale $C$ is denoted by ${\rm Cyl}({\cal A}_C)$ and consists of functions depending on the configurations of finitely many cells. 

The effective theory at scale $C$ is defined after we specify a measure in the space of effective histories that let us calculate the expectation value of physical observables, $\langle f \rangle_C =   \int_{{\cal A}_C} f d\mu_C$. Thus an effective theory at scale $C$  is a pair $({\cal A}_C , \mu_C )$. As mentioned above, in this work we define effective theories for spin systems, sigma models and scalar fields; for gauge systems the different aspects are treated carefully in \cite{GeoC-flat}. 

Given two scales $C_1 \leq C_2$ there are two ways to relate the corresponding effective theories. One map $i_{C_1,C_2}$ in the direction of refinement (which will be shown to induce regularization), and a coarse graining map $\pi_{C_2,C_1}$ 
\[
({\cal A}_{C_1} , \mu_{C_1})  
\begin{array}
{c}
\stackrel{i_{_{C_1 C_2}}}{\longrightarrow} \\ 
\stackrel{\longleftarrow}{\pi_{_{C_2 C_1}}} 
\end{array}
({\cal A}_{C_2} , \mu_{C_2}) .
\]
For these maps $i_{_{C_1 C_2}} \circ \pi_{_{C_2 C_1}}$ is a projection map and 
$\pi_{_{C_2 C_1}} \circ i_{_{C_1 C_2}} = {\rm id}$. 
First let us describe the map in the direction of refinement. Since ${\cal A}_{C_1} \subset {\cal A}_{C_2}$ the needed map is the inclusion map $i_{C_1,C_2} : {\cal A}_{C_1} \to {\cal A}_{C_2}$. 
It can be useful to note that in terms of the spaces ${\cal A}_{L(C)}$ the map in the direction of refinement is 
$i_{C_1,C_2} = r_{2,1}^\ast : {\cal A}_{L(C_1)} \to {\cal A}_{L(C_2)}$. 
Clearly also ${\cal A}_C \subset \bar{\cal A}_M$ and we also have the inclusion map 
$i_C : {\cal A}_C \to \bar{\cal A}_M$. 
This map induces a {\em regularization map} that brings any observable of the continuum to scale $C$, 
$i_C^\ast : {\rm Cyl}(\bar{\cal A}_M) \to {\rm Cyl}({\cal A}_C)$. 
These regularization maps link all the effective theories to the theory at the continuum. 
Thanks to them all the effective theories describe the same physical observables. 
There are many observables of the continuum that get regularized to the same observable at scale $C$. However 
due to the infinite refinement property ($iii$) of our families of cellular decompositions, 
two cylindrical functions of the continuum $f, g \in {\rm Cyl}(\bar{\cal A}_M)$ are different if and only if there is a sufficiently fine scale $C_0$ such that $i_{C_0}^\ast f \neq i_{C_0}^\ast g$. 
This property of regularization on our families of effective configurations says that the subset of $\bar{\cal A}_M$ consisting of elements that are eventually in ${\cal A}_C$ (with respect to the directed partial order of the family) is a dense subset of $\bar{\cal A}_M$. This result can be written symbolically as 
\[
\lim_{C \to M} {\cal A}_C = \bar{\cal A}_M ,
\]
and its main significance for the rest of this work is that it 
will allow us to construct interesting measures in $\bar{\cal A}_M$ as a continuum limit of effective measures.

The coarse graining map $\pi_{_{C_2 C_1}}$ 
models an ``average procedure" by means of decimation. 
For each cell $c_\alpha$ in $C_1$ we choose one ``preferred cell" $c_\alpha'$ in $C_2$ among the cells that intersect $c_\alpha$ 
and define the coarse graining map based on this choice. 
If we are given $s \in {\cal A}_{C_2}$ the configuration 
$\pi_{_{C_2 C_1}} s$ is the $C_1$-constant configuration whose value at $p\in c_\alpha$ is $s(p')$ for any point $p' \in c_\alpha'$. We could say that 
$\pi_{_{C_2 C_1}}$ forgets everything about the configuration $s \in {\cal A}_{C_2}$ except for its values at the ``preferred" cells. 
It is convenient to formulate this definition in terms of the spaces ${\cal A}_{L(C)}$. 
Given a choice of representative embedding 
${\rm Emb}_{1,2}: L(C_1) \to L(C_2)$ our definition is 
$\pi_{_{C_2 C_1}} \doteq {\rm Emb}_{1,2}^\ast : {\cal A}_{L(C_2)} \to {\cal A}_{L(C_1)}$.  
Coarse graining from the continuum is done similarly; after the choice of a 
${\rm Emb}_{L(C)}: L(C) \to M$ we define 
$\pi_{C} \doteq {\rm Emb}_{L(C)}^\ast : \bar{\cal A}_M \to {\cal A}_{L(C)}$. 

Coarse graining maps naturally act on measures letting us calculate expectation values of functions of coarse observables according to the effective theory defined at the fine scale. As in any decimation we simply 
integrate out the degrees of freedom that do not have an impact at the coarser scale, 
\be \label{coarsegrain}
\langle f \rangle_{{C_1}({C_2})}= 
\int_{{\cal A}_{C_1}} f 
( \pi_{_{C_2 C_1 \ast}} d\mu_{C_2}) \doteq 
\int_{{\cal A}_{C_2}} \pi_{_{C_2 C_1}}^\ast f d\mu_{C_2}= 
\langle  (f \circ \pi_{_{C_2 C_1}})\rangle_{C_2} . 
\ee
Similarly, any measure $\mu _M$ 
on $\bar{\cal A}_M$ can be coarse grained to act on effective observables at scale $C$. 
The result of coarse graining is a measure in ${\cal A}_C$ denoted by 
$\mu_C^{\rm ren}=\pi_{C \ast} \mu _M$, 
$\langle f \rangle_C^{\rm ren}= 
\int_{{\cal A}_{C}} f d\mu_C^{\rm ren} \doteq 
\int_{\bar{\cal A}_M}\pi_C^\ast f d\mu _M$. 
The measure $\mu_C^{\rm ren}$ 
gives the correct expectation value according to the theory in the continuum. 

We have defined effective theories, regularization and coarse graining. 
With these elements at hand we now present our {\em implementation of Wilson's renormalization group and the continuum limit}. 

Consider a sequence of increasingly finer scales $\{ C_i \}$. 
If we think 
of the observables of the effective theory as regularizations of observables of the continuum (regularized by the map $i_C^\ast$ defined above), 
we see that the kinematics of the effective theories at any scale describe the same physical system. 
Any two effective theories of the sequence are related  by a coarse graining map, and we have to fix one coarse graining map for each pair of scales of the sequence. 
It is not difficult to see that there are choices of coarse graining maps that are compatible in the sense that coarse graining step by step is equivalent to coarse graining in a single stroke, 
e.g. $\pi_{C_2 C_1} \circ \pi_{C_3 C_2} = \pi_{C_3 C_1}$. 
As mentioned above, these coarse graining maps let us use a more microscopic effective theory to evaluate all the physically important correlation functions and expectation values of a more macroscopic scale (\ref{coarsegrain}). Written in terms of the measures this is 
\be \label{renprescr}
\mu_{C_i}
\approx 
\pi_{C_{i+1} C_i \ast} \mu_{C_{i+1}} ,
\ee
and must be interpreted in one of two alternative frameworks: 
($a$) working on the space of all possible measures or ($b$) working on a truncation that only considers measures of a particular functional form. 

($a$) If we work on the space of all possible measures relation (\ref{renprescr}) is written as an equality and becomes an {\em exact renormalization group transformation}. For any $C_i \leq C_j$ the measure $\mu_{C_j}$
of the effective theory at scale $C_j$ completely determines the expectation values of observables at scale $C_i$ (therefore defining a measure 
$\pi_{C_{j} C_i \ast} \mu_{C_j}$ at scale $C_i$). Clearly in this approach the functional form of the partition function changes under coarse graining. 

($b$) Alternatively we can work at a truncation in which only measures 
written in terms of Boltzman weights 
of a certain functional form are considered. In this framework the measures are labeled by a few coupling constants $\mu_C = \mu_{\beta(C)}$ and the above relation 
(\ref{renprescr}) expresses the intention of 
choosing the measure $\mu_{\beta(C_i)}$ of the coarser theory as the best approximation, among the measures of the desired type, of the coarse graining of the measure $\mu_{\beta(C_{i+1})}$. 

Note that we also consider situations in which the measure $\mu_{\beta(C_{i+1})}$ is not ``homogeneous" in any way and to completely specify it we need {\em local coupling constants} \cite{RGsfRobert}; that is  coupling constants that can take different values at different cells of our cellular decompositions. 
Such a situation arises for example in in-homogenous materials and in systems that do not depend on a metric background. Thus, allowing such systems is essential to include quantum gravity in our framework. 

An equation(s) that formalizes the statement of ``best approximation" (\ref{renprescr}) in terms of asking 
that some physically important correlation functions 
be equal when calculated using the coarse graining of the $C_{i+1}$ effective theory 
or directly the $C_i$ effective theory is called a {\em renormalization prescription}. This equation(s) is solved to determine the coupling constant(s) $\beta(C_i)$ in terms of $\beta(C_{i+1})$ . 
In the general case when there is no homogeneity and we use local coupling constants to specify the measure we need accordingly local renormalization prescriptions \cite{RGsfRobert}. 

When the truncation is appropriately chosen, 
the renormalization group transformation generated by the solving the renormalization prescription is invertible. 
Since our main concern in the rest of this work is constructing a continuum limit,  we will restrict to such cases and 
consider a flow towards the continuum generated by the inverse of the renormalization group transformation%
\footnote{
Strictly speaking, we should in general not talk about a ``renormalization group", but rather about a ``renormalization groupoid" as the group of scale transformations on a metric background is replaced by the groupoid of changes of cellular decomposition (see \cite{RGsfRobert}). 
}. 

When the description of physical observables in terms of effective theories improves as the cut-off scale is refined, 
our collection of effective theories defines a theory in the continuum. 
More precisely, 
consider any observable of the continuum $f \in {\rm Cyl}(\bar{\cal A}_M)$ and 
calculate the expectation value of its 
regularization to scale $C$, $\langle i_C^\ast f \rangle_C$; 
our effective theories 
define a theory in the continuum 
only if these expectation values 
converge as the scale $C$ gets finer and finer. 
That the expectation values converge to $\langle f \rangle_M$, 
\be \label{measureM}
\langle f \rangle_M = \lim_{C \to M} \langle i_C^\ast f \rangle_C \quad , 
\ee
means that given any $\epsilon > 0$ there is a sufficiently fine cellular decomposition $C_0$ such that 
$| \langle f \rangle_M - \langle i_C^\ast f \rangle_C | \leq \epsilon$ for any 
$C\geq C_0$. 
To complete the formal definition 
of the continuum limit 
we only have to say that it is taken in the subfamily of cellular decompositions that are generic with respect to $f$. If the cylindrical function $f$ is sensitive to the collection of $G$-configurations $\{ s(p_1), ... , s(p_n) \}$, the condition on the cellular decompositions is that the points $p_i$ lie on cells of maximal dimension (see section \ref{prelim}). If the limit in equation (\ref{measureM}) exists for every cylindrical function we have given a constructive definition of 
a functional $\mu_M$ in $\bar{\cal A}_M$. The proof of linearity is trivial. The positivity of the $\mu_C$ implies non negativity of $\mu_M$. 

If we have chosen to work with 
a family of cellular decompositions that remains  invariant under the transformations of the group of spacetime symmetries of our system, then no background structure foreign to the symmetry group was used in the construction of the measure (\ref{measureM}). 
Alternatively, we can choose a small, more economic, family of cellular decompositions that is not invariant under the group of spacetime symmetries of our system. In this case, our choice of family is completely arbitrary and it is necessary to check if the resulting measure depends on it before the results are trusted. For example, if the system has the rotation group as a symmetry and the economic family breaks rotational invariance by the introduction of preferred coordinate axis one must check that in the continuum limit rotational symmetry is recovered. 

\noindent
{\em Remark}: \\
Notice that our continuum limit is not a projective limit of measures. The definition relies on the regularization maps $i_C^\ast : {\rm Cyl}(\bar{\cal A}_M) \to {\rm Cyl}({\cal A}_C)$ that are completely novel in loop quantization. The only arrows in the structure of loop quantization before this work were projection maps that could be considered somehow analogous to our coarse graining maps which when acting on functions go in the opposite direction 
$\pi_{C}^\ast :  {\rm Cyl}({\cal A}_C) \to {\rm Cyl}(\bar{\cal A}_M)$.

\section{Explicit example: 2d Ising field theory}

In this case the space of Euclidean histories is the space of spin fields on $\R^2$, 
$\bar{\cal A}_{\R^2}$. We assume that the system that we are describing has physical correlation length $\xi_{\rm phys}$ and that this data is given to us. 

Our approach will use the economic families of cellular decompositions of $\R^2$ 
composed of {\em lattice-type cellular decompositions}. 
Here we describe them in detail. $C_{m,t}$ is a Cartesian lattice-type cellular decomposition of size $a_m= 1/2^m$ (with the directions of coordinate axis fixed). 
The two dimensional cells of $C_{m,t}$ are squares, its one dimensional cells are horizontal or vertical open segments and its zero dimensional cells are points (that we will call vertices). 
We can describe $C_{m,t}$ by the position of its vertices. 
First label the vertices using a pair of integers, say $I \in \Z$ for the $x$ direction and $J \in \Z$ for the $y$ direction, $v^{IJ}$. Their position is given by
$(v^{IJ}_x, v^{IJ}_y) = ( a_m I ,  a_m J) + (t_x, t_y)$. 
(The parameter $t$ 
``slightly shifts" the whole lattice-type cellular decomposition, translating it rigidly by $t$.) 

For any fixed value of the parameter $t$ and letting $m$ run trough the naturals, 
$\{ C_{m,t} \}_{m\in \N}$ is a family of cellular decompositions that satisfies the properties ($i$)-($iii$) of section \ref{prelim}. 
Also, given any collection of marked points $\{ p_1, \ldots , p_n \}$ there are values of $t$ such that $\{ C_{m,t} \}_{m\in \N}$ is {\em generic} with respect to them. We recall that this means that for any cellular decomposition of the family all the marked points fall inside two dimensional cells, 
$\{ \alpha^{(m)}_1, \ldots , \alpha^{(m)}_n \}$. 

The ``$n$-point" functions 
completely characterize the dynamics of an effective theory and later we will use $n$-point functions to characterize the measure in the continuum. At scale $C_{m,t}$ the ``$n$-point functions" are 
\[
\langle s(\alpha_1) \cdots s(\alpha_n) \rangle_{C_{m,t}} =  
\frac{1}{Z_{C_{m,t}}} \sum_s 
\frac{s(\alpha_1) \cdots s(\alpha_n)}{M(\beta_{C_{m,t}})^n} 
\exp{[-\beta_{C_{m,t}} \sum_{(\alpha_j \alpha_k)} s(\alpha_j) s(\alpha_k)]}
\]
where the sum runs over pairs $(\alpha_j \alpha_k)$ of neighboring 2d cells,  $s(\alpha_j)= s(p)$ for any point $p$ in the 2d cell $\alpha_j$, and 
$M(\beta_{C_{m,t}}) = |1 - [\sinh^{-4}(2 \beta_{C_{m,t}})] |^{1/8}$. 

The regularization of the product of $n$ spins at different points of the continuum is simply 
$
i_{C_{m,t}}^\ast ( s({p_1}) \cdots s({p_n})) = 
s(\alpha^{(m)}_1) \cdots s(\alpha^{(m)}_n))
$
where $p_j \in \alpha^{(m)}_j$. 
Thus, after regularization to scale $C_{m,t}$, the $n$-point functions are exactly of the type written above. 

Coarse graining is done 
by a decimation of half of the rows and half of the columns (say keeping the even rows and the even columns). Thus, a single step coarse graining map is defined as 
$
\pi_{m+1,m} = {\rm Emb}_{m, m+1}^\ast
$
with the representative embedding 
${\rm Emb}_{m, m+1}: L(C_{m,t}) \to L(C_{m+1,t})$ defined below. 
To specify the embedding we label each two dimensional cell in $C_{m,t}$ by a pair of integers, 
$\alpha^{(m)}(X , Y)$ corresponding to their $x$ and $y$ coordinates. Then 
${\rm Emb}_{m, m+1}(\alpha^{(m)}(X,Y)) \doteq \alpha^{(m+1)}(2X,2Y)$. 
To complete the prescription of a representative embedding of $ L(C_{m,t})$ we also prescribe the embedding of the one dimensional cells. There are only two choices to embed a one dimensional cell in a representative way: the even and the odd. We again choose the even option. Since we will calculate the continuum limit of 
$n$-point functions using only generic cellular decompositions, the choice of embedding of one dimensional cells turns out to be irrelevant. 

Then the coarse graining of $n$-point functions is simply 
\ba \label{coarsegr}
&&\langle
s(\alpha_1^{(m)}) \cdots s(\alpha_n^{(m)})
\rangle_{C_{m,t}(C_{m+1,t})}= \nl
&&\langle 
s({\rm Emb}_{m, m+1}(\alpha_1^{(m)})) \cdots 
s({\rm Emb}_{m, m+1}(\alpha_n^{(m)}))
\rangle_{C_{m,t}(C_{m+1,t})} .
\ea

Now we will construct a measure in the continuum 
by calculating the continuum limit of the $n$-point functions calculated using the coupling constants that solve an appropriate renormalization prescription. 

Let $\{ p_1, \ldots , p_n \}$ be a set of marked points in $\R^2$. At the scale 
$C_{m,t}$ they induce a set of marked 2-cells $\{ \alpha^{(m)}_1, \ldots , \alpha^{(m)}_n \}$. 
The relative positions of these 2-cells are described in terms of the differences between their $x$ and $y$ coordinates. 
The $x$ coordinate of cell $\alpha^{(m)}_j$  will be denoted by $X_j(m)$ and a similar notation will be used for the $y$ coordinate. Then the integers 
$X_{jk}(m)  \doteq X_j(m) - X_k(m)$, $Y_{jk}(m)  \doteq Y_j(m) - Y_k(m)$ measure the relative position of cells $\alpha^{(m)}_j$, $\alpha^{(m)}_k$ in the lattice of 2-cells. 
Clearly, as we refine the scale 
$C_{m,t} \to \R^2$, $X_{jk}(m) 
\to \infty$, $Y_{jk}(m)\to \infty$ and the size of the cells shrink to zero, $a_m \to 0$. 
However, since the $n$ physical points 
$\{ p_1, \ldots , p_n \}$ are fixed, $a_m X_{\alpha \beta}(m)$ and $a_m Y_{\alpha \beta}(m)$ have a well defined limit. 
At the same time if the coupling constant is properly adjusted, 
the correlation length calculated in lattice units 
$\xi_m$ diverges in a way that should make
$a_m\xi_m$ converge to the physical correlation length of the system $\xi_{\rm phys}$ as $m \to \infty$. 
Recall that by definition the correlation length $\xi$ measures the asymptotic behavior of the correlation function, 
$\langle s(\alpha(0,0)) \cdot  s(\alpha(R,0)) \rangle 
\sim R^{-p}\exp(-R/\xi)$ for large $R$. 
We will postulate as renormalization prescription that 
\be \label{renprescrIsing}
a_m \xi_m = a_m \xi_{m(m+1)} 
\ee
where $\xi_{m(m+1)}$ is the correlation length calculated using the 2-point function 
$\langle s(\alpha_1^{(m)}) \cdot s(\alpha_2^{(m)})
\rangle_{C_{m,t}(C_{m+1,t})}$. 
Notice that, since the correlation length is sensitive only to asymptotic behavior, we have  
$a_m \xi_{m(m+1)} = a_{m+1}\xi_{m+1}$. Thus the renormalization prescription can be written as $a_m \xi_m = a_{m+1}\xi_{m+1}$, or more precisely 
$a_m \xi_m = a_m |z_m^2 +2z_m - 1|^{-1} [z_m(1-z_m^2)]^{1/2}=  \xi_{\rm phys}$ where $z_m= \tanh(\beta_{C_{m,t}})$.%
\footnote{
Another natural renormalization prescription is to ask that a given 2-point function stays fixed after coarse graining. The resulting continuum limit is the same for both renormalization prescriptions. 
} 

In fact, McCoy, Tracy and Wu find convergence in this scaling limit of $n$-point functions of the two dimensional Ising model. In the language of this work, they prove the following theorem \cite{McCoyetal}. 
\begin{theorem}[McCoy, Tracy, Wu]
Choose $\beta_{C_{m,t}}$ as to satisfy the renormalization prescription (\ref{renprescrIsing}). Then 
a measure $\mu_{\R^2}$ in $\bar{\cal A}_{\R^2}$ is defined by its $n$-point functions that are calculated as a continuum limit 
\[
\lim_{{C_{m,t}} \to \R^2} 
\langle s(p_1) \cdots s(p_n) \rangle_{C_{m,t}} = 
\langle s(p_1) \cdots s(p_n) \rangle_{\R^2} . 
\]
Moreover, explicit expressions for these $n$-point functions are given below. 
\end{theorem}
In the case that the coupling constants involved in the renormalization group flow are below the critical point, $\beta_{C_{m,t}} \leq \beta_{\rm critical}= \tanh^{-1}(\sqrt{2} -1)$ they show: 
\[
\lim_{{C_{m,t}} \to \R^2} 
\langle s(p_1) \cdots s(p_n) \rangle_{C_{m,t}} = 
\exp\Big(\sum_{k=2}^{\infty}f_n^{(k)}\Big) 
\]
where
\[
f_n^{(k)}= \frac{-1}{2k (2\pi^2)^k}\!\!\int_{-\infty}^{\infty}\!\!\!\!\textrm{d}v_1\cdots\textrm{d}v_k\textrm{d}u_1\cdots\textrm{d}u_k
\prod_{l=1}^{k}\Big(1+v_l^2+u_l^2 \Big)^{-1}\frac{u_l+u_{l+1}}{v_l-v_{l+1}+i\epsilon}\textrm{Tr}[\prod_{r=1}^{k}A(r)]
\]
and the entries of the $n \times n$ matrix $A(r)$ are 
$A(r)_{ij}=\textrm{sgn}(x_{jk}) \exp(-i x_{ij}y_r - i y_{ij} x_r)$, where 
$x_{\alpha \beta}= X_{\alpha \beta}/ \xi_m$ and 
$y_{\alpha \beta}= Y_{\alpha \beta}/\xi_m$. The diagonal matrix elements of $a(r)$ vanish. The explicit form of the correlation length is 
$\xi_m = |z_m^2 +2z_m - 1|^{-1} [z_m(1-z_m^2)]^{1/2}$ for $z_m= \tanh(\beta_{C_{m,t}})$. 

In the case that the coupling constants involved satisfy $\beta_{C_{m,t}} \geq \beta_{\rm critical}$ they show: 

\[
\lim_{{C_{m,t}} \to \R^2} 
\langle s(p_1) \cdots s(p_n) \rangle_{C \beta_{C_{m,t}}} = 
\vert\textrm{det}(\sum_{k=1}^{\infty}g_{(n)ij}^{(k)})\vert^{1/2}
\exp\Big(\sum_{k=2}^{\infty}f_n^{(k)}\Big) 
\]
where
\[
g_{(n)ij}^{(k)}=\frac{1}{(2\pi^2)^k}\!\!\int_{-\infty}^{\infty}\!\!\!\!\textrm{d}y_1\cdots\textrm{d}y_k\textrm{d}x_1\cdots\textrm{d}x_k
\prod_{l=1}^{k}\Big(1+x_l^2+y_l^2 \Big)^{-1}\prod_{l=1}^{k-1}\frac{y_l+y_{l+1}}{x_l-x_{l+1}+i\epsilon}[\prod_{r=1}^{k}A(r)]_{ij} . 
\]

Once we have constructed the theory in the continuum we can go back to scale $C_{m,t}$ and calculate corrections to the effective theory. The completely renormalized theory at scale $C_{m,t}$ 
is determined by the coarse graining of $n$-point functions from the continuum
\[
\langle  s(\alpha_1) \cdots s(\alpha_n) 
\rangle_{C_{m+1,t}}^{\rm ren}= 
\langle  s(p_1) \cdots s(p_n) 
\rangle_{\R^2} 
\]
with $p_j= {\rm Emb}_{L(C_{m,t})} \alpha_j$. 

The scaling limit of the 2d Ising model has been extensively studied (for a review see \cite{McCoyRev}). Of particular importance is the proof \cite{OS2dIsing} showing 
that the $n$-point functions in the continuum defined above satisfy the Osterwalder-Schrader axioms \cite{OS}. In particular the rotational invariance that was broken by working with an economic family is known to be restored in the continuum limit. 
Regarding the implied relativistic quantum field theory, 
we can explicitly give a covariant Hamiltonian quantum theory following an Osterwalder-Schrader type construction on the space $\bar{\cal A}_M$ like the one proposed in \cite{ALMMT}. Explicit knowledge of the relation between this quantum field theory and the standard realization of the 2d Ising field theory [$\phi^4$ Landau-Ginzburg model] 
in terms of linear scalar fields would certainly be desirable. 
If we knew such a relation we could import very important concepts into the loop quantized world. For example, 
for the standard realization a basis of quasi particles has been found and in that basis the $S$-matrix is explicitly known \cite{S-matrix}.

We would like to thank M. Reisenberger, H. Sahlmann, F. Markopoulou, V. Husain, A. Corichi, A. Ashtekar and A. Perez for enlightening conversations at different stages of this project. 
This work was partially supported by CONACyT grants 40035-F and U47857-F and  DGAPA grant IN108803-3. 


\end{document}